
\magnification=\magstephalf
\nopagenumbers
\parindent=0pt
\parskip=5pt
\hsize=16 true cm
\vsize=25.5 true cm
\voffset=-0.2 truein
\font\rfont=cmr9 at 9 true pt
\def\ref#1{$^{\hbox{\rfont {[#1]}}}$}


\catcode`\@=11 

\def\nolabels{\def\wrlabeL##1{}\def\eqlabeL##1{}\def\reflabeL##1{}}
\def\writelabels{\def\wrlabeL##1{\leavevmode\vadjust{\rlap{\smash%
{\line{{\escapechar=` \hfill\rlap{\sevenrm\hskip.03in\string##1}}}}}}}%
\def\eqlabeL##1{{\escapechar-1\rlap{\sevenrm\hskip.05in\string##1}}}%
\def\reflabeL##1{\noexpand\llap{\noexpand\sevenrm\string\string\string##1}}}
\nolabels
\global\newcount\refno \global\refno=1
\newwrite\rfile
\def\defref{$^{{\hbox{\rfont [\the\refno]}}}$\nref}
\def\nref#1{\xdef#1{\the\refno}\writedef{#1\leftbracket#1}%
\ifnum\refno=1\immediate\openout\rfile=refs.tmp\fi
\global\advance\refno by1\chardef\wfile=\rfile\immediate
\write\rfile{\noexpand\item{#1\ }\reflabeL{#1\hskip.31in}\pctsign}\findarg}
\def\findarg#1#{\begingroup\obeylines\newlinechar=`\^^M\pass@rg}
{\obeylines\gdef\pass@rg#1{\writ@line\relax #1^^M\hbox{}^^M}%
\gdef\writ@line#1^^M{\expandafter\toks0\expandafter{\striprel@x #1}%
\edef\next{\the\toks0}\ifx\next\em@rk\let\next=\endgroup\else\ifx\next\empty%
\else\immediate\write\wfile{\the\toks0}\fi\let\next=\writ@line\fi\next\relax}}
\def\striprel@x#1{} \def\em@rk{\hbox{}}
\def\lref{\begingroup\obeylines\lr@f}
\def\lr@f#1#2{\gdef#1{\defref#1{#2}}\endgroup\unskip}
\newwrite\lfile
{\escapechar-1\xdef\pctsign{\string\%}\xdef\leftbracket{\string\{}
\xdef\rightbracket{\string\}}}

\def\writestop{\def\writestoppt{\immediate\write\lfile{\string\p
ageno%
\the\pageno\string\startrefs\leftbracket\the\refno\rightbracket%
\string\def\string\secsym\leftbracket\secsym\rightbracket%
\string\secno\the\secno\string\meqno\the\meqno}\immediate\closeout\lfile}}
\def\writestoppt{}\def\writedef#1{}
\catcode`\@=12 

\centerline{SMALL $x$ AND SMALL $Q^2$}
\bigskip
\centerline{P.\ V.\ Landshoff}
\centerline{DAMTP, University of Cambridge}
\vskip 20pt
\leftline{{\bf Abstract}}
\medbreak
I discuss (i) hadron-hadron and photon-hadron
total cross-sections, (ii) elastic scattering,
(iii) the r\^{o}le of minijets in total
cross-sections and the need to include them in
Monte-Carlos for small-$x$ structure functions,
(iv) how the two pomerons affect small-$x$
behaviour, and (v) rapidity-gap physics.
\vskip 20pt
\leftline{1. {\bf Total cross-sections}}
\medbreak
This talk is about nonperturbative effects at
high energy, and this means that its central
theme is Regge theory\defref\pdb{
P D B Collins, {\it Introduction to Regge theory}, Cambridge University
Press (1977)
}.  I remind you that if one
plots the spins $\alpha$ of families of particles
against their squared masses $t$, one finds
linear behaviour. The best example is the $\rho,
\omega, f, a$ families, shown in figure 1.  Regge
theory tells us that if we extrapolate the line
$\alpha (t)$ to negative $t$, so that $t$ may
then be interpreted as a momentum transfer, the
sum of the exchanges of all the particles in any
of the families contributes to a high-energy
elastic scattering process at centre-of-mass
energy $\sqrt{s}$ and momentum transfer $t$ as
follows:
$$T(s,t)\sim\beta(t)\,\xi_{\alpha}(t)\,s^{\alpha(t)}\eqno{(1)}$$
Here $\beta(t)$ is some unknown real function and
$\xi_{\alpha}(t)$ is a definite phase calculated
from $\alpha (t)$.  We call such a sum of
exchanges Reggeon exchange, and depict it
diagrammatically by a jagged line as in figure 2.

The optical theorem allows us to calculate the
total cross-section from the elastic amplitude:
$$\sigma^{{\rm TOT}}={1\over s}\,T(s,t=0)\sim
s^{\alpha(0)-1}.\eqno{(2)}$$
According to figure 1, for the $\rho,\omega,f,a$
families $\alpha (0) \approx {1\over 2}$, so that
their contribution to $\sigma^{{\rm TOT}}$ has
behaviour close to $1/\sqrt{s}$.  In order to
obtain rising total cross-sections, as are found
experimentally, we need to introduce another
Reggeon, known as the soft pomeron.  This has
$$\alpha(0)=1+\epsilon\qquad
(\epsilon>0)\eqno{(3)}$$ and in order to describe
the data we need\defref\08{
P D B Collins and F Gault, Physics Letters 112B (1982) 255}\defref\elastic{
A Donnachie and  P V Landshoff, Nuclear Physics B267 (1986) 690
} $\epsilon\approx$ 0.08: see
figure 3.\hfil\break

The soft pomeron is nonperturbative in origin,
and is almost certainly a consequence of
nonperturbative gluon exchange\defref\gluon{
P V Landshoff and O Nachtmann, Z Physik C35 (1987) 405;
F Halzen, G Krein and A A Natale, Physical Review D47 (1993) 295;
J R Cudell and B U Nguyen, McGill preprint 93-25
}.  It would be nice
to think that actually it is associated with
glueball exchange, so that one can make a plot
something like figure 1 but with glueballs
instead of quark states.  As I shall explain,
near $t = 0$ the slope of the pomeron trajectory
is somewhat smaller than for $\rho,\omega,f,a$:
$$\alpha(t)=1+\epsilon+\alpha't\eqno{(4)}$$
with $\alpha'=$ 0.25 GeV${}^{-2}$ instead of the
$0.8 =$ GeV${}^{-2}$.  If $\alpha (t)$ remains
straight for positive $t$, there should be a
$2^{++}$ glueball with mass about $1900$ MeV.  If
glueballs are not found, then pomeron exchange is
probably just the exchange of a number of
separate nonperturbative gluons.

The other pomeron, the Lipatov pomeron\defref\lipatov{
E A Kuraev, L N Lipatov and V Fadin, Soviet Physics JETP 45 (1977) 199;
},
corresponds to perturbative gluon exchange.  It
has a significantly larger value of $\epsilon$,
about $0.5$.  There is no firm evidence that it
has yet been seen in any experiment, though there
is the hope that the HERA measurements of $\nu
W_2$ at very small $x$ may at last have revealed it.
Maybe, just maybe, it has even been seen in the
$\bar{p}p$ total cross-section measured at the
Tevatron\defref\Margolis{
J R Cudell and B margolis, Physics Letters B297 (1992) 398
}.  In figure 3, I show the published
$\sqrt{s} = 1800$ GeV measurement of the E170
experiment.  The CDF experiment this summer has
reported a much larger value, $\sigma \bar{p}p =80.6\pm2.3$
mb.  If it should turn out that the conflict
between these two results is resolved in favour
of CDF, it might indicate the onset of
Lipatov-pomeron exchange.

\vskip 20pt
\leftline{2. {\bf Elastic Scattering}}
\medbreak
The form (1) may be compared with
elastic-scattering data.  Long ago\defref\jaros{
G A Jaroskiewicz and P V Landshoff, Physical Review D10 (1974) 170
},  the ISR $pp$
data yielded the value $\alpha' = 0.25$, and this
then led to remarkably successful predictions\ref{\elastic} for
$\bar{p}p$ elastic scattering at the CERN
collider and at the Tevatron: see figure 4.  This
is a real triumph for Regge theory: between ISR
and Tevatron energies the change in the forward
elastic slope is more than $3$ GeV$ ^{-2}$ and is
a direct confirmation that the value of $\alpha'$
is correct.  Likewise, the phase of the forward
amplitude is well predicted, now that the UA4
collaboration has changed its result at CERN
collider energy\defref\Ua4{
UA4 collaboration, Physics Letters B316 (1993) 448
}.

I should mention that we know, from unitarity,
that it is not enough to consider just the
exchange of a single pomeron.  One must add in
also the simultaneous exchange of two or more
pomerons.  Even after more than 30 years of Regge
theory, we do not have any certain way to
calculate these higher exchanges.  Unlike other
authors, Donnachie and I believe\ref{\elastic} that their
effect at present energies is small in the
forward direction (less than 10\%), though they
are essential away from $t=0$.

\vskip 20pt
\leftline{3. {\bf Minijets}}
\medbreak
I have shown the successful prediction from Regge
theory for the $\gamma p$ total cross-section at
HERA.  There were some other predictions that
were much larger, and it is interesting to
discuss why they turned out to be wrong.

They were mostly based on perturbative
calculations of inclusive minijet production.
Since the scale of nonperturbative effects is
typically $1$ GeV, I would expect that
perturbation theory reproduces the inclusive
cross-section $d\sigma/dp_T$ reasonably
successfully (say, to within better than a factor
of $2$) down to $p_T^{{\rm min}}
\approx 1$ GeV.

Now\defref\jacob{
M Jacob and P V Landshoff, Mod Phys Lett A1 (1986) 657
}
$$\int_{p_T^{{\rm min}}}dp_T\,{d\sigma\over dp_T}=\langle
n\rangle\;\sigma (p_T>p_T^{{\rm min}})\eqno{(5)}$$
where $\sigma\left(p_T>p_T^{{\rm min}}\right)$ is
the contribution to $\sigma^{{\rm TOT}}$ from
events containing minijets, and $\langle
n\rangle$ is the average number of minijets in
these events.  One calculates $d\sigma/dp_{T}$
from the familiar diagram of figure 5a, which
involves the structure functions of the incoming
particles and a hard scattering.  The naive
expectation from figure 5a is that exactly 2
minijets are produced, that is $\langle n\rangle
= 2$ in (5).  However, if one calculates the
left-hand side of (5) and compares with the HERA
measurements, this would require $\sigma
(p_T>p_T^{{\rm min}}) >
\sigma^{{\rm TOT}}$, which is wrong by definition.

So it must be that $\langle n\rangle > 2$.  This
is not really a surprise: it has long been known
that it is correct to use figure 5a to
calculate the inclusive cross section
$d\sigma/dp_T$, but\defref\detar{
C E DeTar, S D Ellis and P V Landshoff, Nuclear Physics B87 (1975) 176
} that this simple figure
cannot be expected correctly to reproduce other
features of the event structure.  One way in
which $\langle n\rangle$ can be greater than 2
is from multiple parton-parton scatterings
involving two or more partons from each of the
initial particles.  I would expect this to be
very important in high-energy nucleus-nucleus
reactions\defref\kajantie{
K Kajantie, P V Landshoff and J Lindfors, Physical Review Letters
59 (1987) 2527
}, but I doubt whether it is the main
mechanism in $\gamma p$ collisions.

Rather, there is another mechanism\ref{\jacob} which is
intrinsically nonperturbative and which has so
far not been included in the Monte Carlos that
are used for HERA data analysis.  In order to
calculate $d\sigma/dp_T$ from figure 5a, one needs
the two structure functions down to
fractional-momentum values $x$ of order $x \sim
p_T^{{\rm min}}/\sqrt{s}$, that is very small
values when $\sqrt {s}$ is large.  But, from
elementary kinematics, when a parton of momentum
$k$ is pulled out of a particle of momentum $p$ with
$k = xp + \dots\;$, the squared invariant mass of
system it leaves behind is\defref\short{
P V Landshoff, J C Polkinghorne and R D Short, Nuclear Physics
B28 (1970) 210
}
$$s_0\sim - {k^2+{\bf k}^2_T\over x}\eqno{(6)}$$
and so is large when $x$ is small.  That is, the
upper and lower clusters of residual fragments of
the initial hadrons in figure 5a each have large
invariant mass and so they are very likely to
contain additional minijets.

Another way to understand this is from a ladder
model for the total cross-section: figure 5b.  Of
course, one must integrate over the transverse
momentum round each loop of the ladder.  Consider
the contribution from a ``hot'' loop somewhere in
the middle of the ladder, that is from large
transverse momentum in that loop.  If I cut the
ladder down the middle and draw circles round the
lines above and below the hot loop, as I have
done in the figure, the left-hand side of the
diagram has exactly the form of figure 5a.
(Including also the right-hand side just squares
it, as is needed to calculate the cross-section.)
However, it may well be that another loop of the
diagram is hot too, or even several other loops,
corresponding to several pairs of minijets.  As
is well known\ref{\lipatov}, at small $x$ there is no $k_T$
ordering, so the hot loops are most likely not
next to one another, but rather are separated by
loops that are not hot, so that one cannot
calculate the multi-minijet production purely
from perturbation theory.

Consider now $\nu W_2$ at a value of $Q^2$ that
is only moderately large, so that perturbative
evolution has not yet set in and the parton model
applies: figure 6.  The squared energy of the
lower bubble, which is the amplitude for finding
a parton $k$ in the proton $p$, again satisfies
(6) and so is large at small $x$.  Since this bubble is
an elastic strong-interaction amplitude, its high-energy
behaviour is governed by Regge theory and is just
a sum of terms $s_0^{\alpha(0)}$.  If we insert
this in the calculation of $\nu W_2$ from figure
7, we obtain\ref{\short} a sum of terms $(1/x)^{\alpha (0) -
1}$.

It is well known that $\nu W_2$ contains such
Regge terms at small $x$, approximately a
constant from soft-pomeron exchange and close to
$\sqrt{x}$ from $f, a$ exchange.  What is not so
well known is that such a behaviour, which I
stress is nonperturbative, arises because the
proton fragments that remain when a small-$x$
parton is pulled out have large invariant mass.

The Monte Carlos used at HERA set both $k^2$ and
$k_T$ to zero (before the perturbative evolution
begins) and so do not make $s_0$ large.  Thus
they miss a nonperturbative effect which is
surely important.

\vskip 20pt
\leftline{4. {\bf The two pomerons}}
\medbreak
As I have explained, at least at moderate $Q^2$
the small-$x$ behaviour of $\nu W_2$ should
contain the same powers of $1/x$ as the powers of
$s$ that appear in the fit of figure 3 to the
$\gamma p$ total cross section.  Another
important feature of $\nu W_2$ is that, at $Q^2 =
0$, it vanishes linearly with $Q^2$.  Indeed,
$$\sigma^{\gamma p}={4\pi^2\alpha\over Q^2}\
\nu\,W_2\Big\vert _{Q^2=0}.\eqno{(7)}$$
The simplest fit to the small-$x$ data that has these
features is provided for by the form
$$\nu W_2=X\,x^{-0.0808}\,\left(Q^2\over Q^2+a^2\right)^{1.0808}+
Y\,x^{0.4525}\,\left(Q^2\over Q^2+b^2\right)^{0.5475}.\eqno{(8)}$$
Figure 7 shows the result of such a fit\defref\smallx{
A Donnachie and P V Landshoff, preprint DAMTP 93-23\ \  M/C-TH 93/11,
to appear in Z Physik C
} to the
small-$x$ NMC data up to $Q^2 = 10 $ GeV$^2$.  It
is a two-parameter fit, because for each choice
of $X$ and $Y$ one determines $a$ and $b$ in such
a way as to retrieve, through (7), the $\gamma p$
fit shown in figure 3. The best-fit values are
$a=750$ MeV and $b=110$ MeV.

If one sets $Q^2 = 8.5$ and extrapolates the fit
down to $x = 2 \times 10^{-4}$, one obtains $\nu
W_2 \approx 0.6$.  This is much smaller than the
measured value\defref\H1{
H1 collaboration, preprint DESY 93-117
} of about $1.4 \pm 0.5$ reported by
H1, which raises the exciting possibility that
the discrepancy may be attributed to the presence
of the second pomeron, the Lipatov pomeron.  In
order to decide whether this is true, it is
necessary to learn how the two pomerons live
together.  There are those who believe that, as
$Q^2$ increases, the soft pomeron goes smoothly
over to the Lipatov pomeron, that is $x^{-0.08}$
changes to something like $x^{-0.5}$.  However, I
think it is more likely that the two terms should
be \underbar{added} together.

This is because of a simple model of the Lipatov
pomeron that Collins and I studied a couple of
years ago\defref\collins{
J C Collins and P V Landshoff, Physics Letters B276 (1992) 196
}.
We wrote  the Lipatov equation\ref{\lipatov} as
$$T=T_0+K\otimes T\eqno{(9)}$$
where the last term denotes, as usual, a
convolution of the amplitude with the Lipatov
kernel $K$ that includes an integration over
transverse momentum $k_T$.  The kernel $K$ is
calculated from perturbation theory, and so it is
not valid to use it below some value $Q_0$ which
is of order $1$ GeV.  So we restricted the
integration to $k_T > Q_0$.  In order to include
nonperturbative effects associated with smaller
values of $k_T$, we chose the driving term $T_0$
in the equation to correspond to soft-pomeron
exchange.  In the generalised ladder-like
diagrams that give the Lipatov equation, this
amounts to restricting the small $k_T$ to the
part of the ladder near to the proton, and taking
all the transverse momenta nearer to the photon
to be perturbative.  This cannot be more than an
approximation --- I have already remarked that, at
small x, there is no $k_T$ ordering --- but it is
better than not including nonpertubative effects
at all.  In order to be able to solve the Lipatov
equation exactly, we approximated the kernel $K$
by a simple function that has the essential
features of the true kernel.  The resulting output
amplitude $T$ was a sum of terms corresponding to
the Lipatov power $x^{-0.5}$ and the input
driving term $x^{-0.08}$

We studied\ref{\collins} another important effect with our
equation, that arising from energy conservation,
by imposing also an upper limit $Q_1$ on the
$k_T$ integration.  Before the imposition of this
upper limit, the $x^{-0.5}$ term is in fact
accompanied by a multiplicative logarithmic
factor.  With the upper limit, it is transformed
into a sum of pure powers $x^{-N}$.  The two
leading values of $N$ are
$$-0.5 (1 - \Delta^2),
\hskip 0.5 in -0.5(1 - 4\Delta^2)$$

 where $\Delta = \left [\pi/\hbox{log}(Q_1/Q_0)
\right ]$.  There are an infinite number of such
powers, and as $Q_1 \rightarrow \infty$ they all
coalesce at $x^{-0.5}$.  However, with $Q_0 = 1$
GeV and $Q_1$ the maximum $\gamma^\ast p$ energy
attainable at HERA, the leading power is only
$x^{-0.35}$.  Thus the energy-conservation
corrections to the Lipatov equation are
substantial.  This has recently been verified in
the framework of the exact Lipatov equation by
Forshaw, Harriman and Sutton\defref\forshaw{
J Forshaw, P N Harriman, and P J Sutton,
preprint RAL-93-039 \ \  M/C-TH-93-14
}, while Bartels and Lotter\defref\bartels{
J Bartels and H Lotter, Physics Letters B309 (1993) 400
} have
confirmed the important effects arising from the
lower limit $Q_0$.

\vskip 20pt
\leftline{5. {\bf Rapidity-gap physics}}
\medbreak
Figure 8 depicts the process known in
proton-proton or proton-antiproton collisions as
diffraction dissociation, in which one of the
incoming hadrons emerges with only very small
change of momentum.  A very small fraction $\xi$,
less than $5\%$ or so, of its initial momentum is
supposed to have been carried off by a pomeron,
which collides with the other hadron.  The cross
section is of the form\defref\structure{
A Donnachie and P V Landshoff, Nuclear Physics B303 (1988)  634
}
$${d^2\sigma\over dt\,d\xi}=F_{P/p}(t,\xi)\,\sigma^{Pp}\eqno{(10)}$$
where $t$ is the momentum transfer from the
initial to the final fast proton and $F_{P/p}$ is
the probability for it to have ``radiated'' a
pomeron:
$$F_{P/p}(t,\xi)={9\beta^2\over
4\pi^2}\,\left[F_1(t)\right]^2\,\xi^{1-2\alpha(t)}\eqno{(11)}$$
with $\beta^2 \approx 3.5$ GeV$^{-2}$ and
$F_1(t)$ the elastic form factor of the proton.
It is not essential to insist that the fast
proton reaches the final state without breaking
up; if one does not, then one should omit the
elastic form factor.  The quantity $\sigma^{Pp}$
is, by definition, the cross section for the
pomeron interacting with the other proton and is
the object of interest.  (It should be noted that
some authors adopt a definition of $F_{P/p}$ that
differs from mine by a factor $\pi/2$, with a
corresponding adjustment to $\sigma ^{Pp}$).

Because there is a very fast proton (whether or
not it has been allowed to break up), there is a
rapidity gap in the final state; that is just
kinematics.  Note, however, that the presence of
the gap is
\underbar {not} sufficient to guarantee that pomeron-exchange is
involved.  There is good evidence that often instead it
is $f, \rho, \dots$ or even a pion.  In order to
check that there is no such contamination, one
needs to check the presence of the factor
$\xi^{1-2\alpha (t)}$ with $\alpha (t)$ the
pomeron trajectory (4), rather than with the $\rho,
\omega, f, a$ trajectory of figure 1 or the pion
trajectory.  In order to minimise the
contamination\defref\diffdis{
A Donnachie and P V Landshoff, Nuclear Physics B244 (1984)  322
},
$\xi$ should be as small as
possible, but $t$ should not be too small (though
not more than about $1$ GeV$^2$, so that one may
reasonably expect that only a single pomeron is
being exchanged).

Some part of the pomeron-proton cross section
$\sigma^{Pp}$ is expected to be contributed from
events with high $p_T$ jets.  That this is so has
been verified by the UA8 experiment at the CERN
collider, which finds\defref\ua{
P Schlein, talk at European HEP Conference, Marseille (1993)
} events with $p_T > 8$ GeV/c
jets.  To calculate the cross-section expected
for such events, one needs to know the pomeron
structure function\defref\ingsch{
G Ingelman and P Schlein, Physics Letters B152 (1985) 256
}.  We know that the pomeron is
exchanged between quarks, since pomeron exchange
generates the total hadron-hadron cross sections
shown in figure 3.  Further, it seems to couple
to single quarks: the evidence for this is that\defref\sigfits{
A Donnachie and P V Landshoff, Physics Letters B296 (1992) 227
} the $s^{0.08}$ term in $\sigma^{{\rm TOT}} (\pi
p)$ is close to $2/3$ its magnitude in
$\sigma^{{\rm TOT}}(pp)$.  Thus, even though we
believe that what is exchanged is gluons, they
are coupling to quarks when they generate the
pomeron.  This led Donnachie and me to suppose\ref{\structure}
that the pomeron structure function is dominantly
quark; unlike most other authors, we believe that
its gluon component is rather small.  Hence our
diagram for the UA8 process is that of figure
10a, where one of the quarks to which the pomeron
couples participates in the hard scattering that
produces the high-$p_T$ jets, and the other is a
longitudinal spectator.  We find that the quark
structure function of the pomeron is, for each
light quark or antiquark,
$$x\,q_P(x)=\textstyle{1\over 3}C\pi x(1-x)\qquad(x\geq\quad 0.1)\eqno{(12)}$$
where $C$ is the coefficient of $x^{-0.08}$ in
the corresponding total proton quark
distribution.  Note that $x$ is the Bjorken fractional-momentum variable
relative to the momentum of the pomeron. The fit to the data shown in
figure 7 corresponds to $C=0.23$, a somewhat
larger value than the value we originally used\ref{\structure},
which came from EMC data.  If we sum over all the
quarks, we obtain $x(1-x)$ with multiplying
coefficient just a little greater than $1$.

The UA8 data\ref{\ua} seem to be in agreement with
this, though it is not possible to determine from
those data whether the pomeron structure function
is quark or gluon.  But the data do rule out the
possibility that the structure function saturates
a momentum sum rule, which was expected by some
authors; as the pomeron is not a state, but only
something that is exchanged, there is no reason
to have saturation.

An interesting feature of the UA8 data is that in
a significant fraction of events all, or nearly
all, the energy of the pomeron-proton collision
is taken by the pair of high-$P_T$ jets.  The
natural explanation\defref\unusual{
A Donnachie and P V Landshoff, Physics Letters B285 (1992)172
} is that these events are
generated not by a process of the type of figure
9a, but rather figure 9b, which has no
longitudinal spectator jet coupled to the
pomeron.

Important information about the pomeron structure
function will come from HERA, since
$\gamma^*$-pomeron collisions probe directly the
quark structure function.  With our structure
function (12), at small $x$ some 10\% of the
total events that make up the total $\nu W_2^p$
should be expected\ref{\structure} to have a very fast proton.
One can also look for the analogue of the events
corresponding to figure 9b, namely fast-proton
events with two high-$P_T$ jets and nothing else.
One can use real photons to look for these, and
the cross-section should be quite large\ref{\unusual},
$\sigma^{\gamma p}\approx 1 nb$ for $p_T > 5$
GeV/c.

\vfill\eject
\leftline{6. {\bf Conclusions}}
\medbreak
Pomeron physics began more than 30 years ago and
is nowadays a very active and interesting area of
study.  The phenomenology of the soft pomeron is
surprisingly simple and has allowed several
successful predictions.  Soft pomeron exchange is
nonperturbative gluon exchange, or maybe glueball
exchange. There is some understanding of the
theory behind this, but this is a difficult
subject and further work on it will need to be
guided by more data, such as will come from HERA.
Finally, the possibility that HERA will discover
also the Lipatov pomeron, which corresponds to
perturbative gluon exchange, opens up a new
opportunity for studying aspects of QCD.

\vskip 20pt
\medskip\immediate\closeout\rfile\writestoppt
\baselineskip=4pt{{\bf References}}\medskip{\frenchspacing%
\parindent=20pt\escapechar=` \input refs.tmp\bigskip}\nonfrenchspacing
\vfill\eject
{}.
\vskip 12 truecm
{Figure 1} The $\rho ,\omega, f,a$ trajectory
{}.
$\phantom{x}$
\vskip 55truemm
{Figure 2} Reggeon exchange. The jagged line represents a sum of particle
exchanges
\vfill\eject
{}.
\vskip 22truecm
{Figure 3} Total cross-sections
\vfill\eject
{}.
\vskip 9truecm
{Figure 4} ISR data for $pp$ and Tevatron data for
$\bar pp$ elastic scattering, with 1985 curves from reference {\elastic}
{}.
\vskip 8truecm
{Figure 5} Mechanism for high-$p_T$  jet production: ($a$)
central hard scattering and two structure functions;
($b$) a ladder model -- the circles
represent the structure functions
{}.
\vskip 50truemm
{Figure 6} The parton model for $\nu W_2$
\vfill\eject
{}.
\vskip 85truemm
{Figure 7} NMC data for $\nu W_2$, with fit of the form (8)
{}.
\vskip 55truemm
{Figure 8} Diffraction dissociation
{}.
\vskip 75truemm
{Figure 9} ($a$) Diffractive high-$p_T$  jet production; ($b$) an
additional mechanism

\bye